\documentclass[twocolumn,amssymb, amsmath, nobibnotes, aps, superscriptaddress,pre]{revtex4-1}   

\usepackage{color, graphicx, hyperref, multirow, url, booktabs, colortbl, makecell, pbox}

\newcommand{\ds}{\displaystyle}

\newcommand{\be}{\begin{enumerate}}
\newcommand{\ee}{\end{enumerate}}

\newcommand{\wrp}[1]{\left( #1 \right)}
\newcommand{\swrp}[1]{\left[ #1 \right]}

\newcommand{\kmin}{k_{\text{min}}}
\newcommand{\ntail}{n_{\text{tail}}}

\begin{document}
\author{Anna D. Broido}%
\email{anna.broido@colorado.edu}
\affiliation{Department of Applied Mathematics, University of Colorado, Boulder, CO, USA}

\author{Aaron Clauset}
\email{aaron.clauset@colorado.edu}
\affiliation{Department of Computer Science, University of Colorado, Boulder, CO, USA}
\affiliation{BioFrontiers Institute, University of Colorado, Boulder, CO, USA}
\affiliation{Santa Fe Institute, Santa Fe, NM, USA}

\title{Scale-free networks are rare}%
\begin{abstract}
A central claim in modern network science is that real-world networks are typically ``scale free," meaning that the fraction of nodes with degree $k$ follows a power law, decaying like $k^{-\alpha}$, often with $2 < \alpha < 3$. However, empirical evidence for this belief derives from a relatively small number of real-world networks. We test the universality of scale-free structure by applying state-of-the-art statistical tools to a large corpus of nearly 1000 network data sets drawn from social, biological, technological, and informational sources. We fit the power-law model to each degree distribution, test its statistical plausibility, and compare it via a likelihood ratio test to alternative, non-scale-free models, e.g., the log-normal. Across domains, we find that scale-free networks are rare, with only 4\% exhibiting the strongest-possible evidence of scale-free structure and 52\% exhibiting the weakest-possible evidence. Furthermore, evidence of scale-free structure is not uniformly distributed across sources:\ social networks are at best weakly scale free, while a handful of technological and biological networks can be called strongly scale free. These results undermine the universality of scale-free networks and reveal that real-world networks exhibit a rich structural diversity that will likely require new ideas and mechanisms to explain.
\end{abstract}

\maketitle

%\section{Introduction}
Networks are a powerful way to both represent and study the structure of complex systems of all kinds. Examples today are plentiful and include social interactions among individuals, both offline and online, protein or gene interactions in biological organisms, communication between digital computers, and various kinds of transportation networks. Across scientific domains and different types of networks, it is common to encounter the claim that most or all real-world networks are \textit{scale free}. The precise details of this claim vary across the literature~\cite{Albert1999, Przulj2007, Lima-Mendez2009, Mislove2007, Agler2016, Ichinose2017, Zhang2015}, but generally agree that a network is scale free if the fraction of nodes with degree $k$ follows a power-law distribution $k^{-\alpha}$, where $\alpha>1$. Some versions of this ``scale-free hypothesis" make the requirements stronger, e.g., requiring that $\alpha \in [2,3]$ or that node degrees evolve by the preferential attachment mechanism~\cite{Dorogovtsev2002, Barabasi1999}.  Other versions make them weaker, e.g., requiring that the power law holds only in the upper tail \cite{Willinger2009} or is merely more plausible than a thin-tailed distribution like an exponential or normal~\cite{Albert2000}.

Despite the frequency of these claims, there has been no broad evaluation of the empirical prevalence of scale-free patterns in real-world networks. If networks are in fact universally scale free, then many theoretical results about scale-free networks have broad scientific relevance. For instance, the several mechanisms known to build scale-free networks~\cite{Price1965, Simon1955, Barabasi1999,Dorogovtsev2002,Pastor-Satorras2003,Berger2004,Leskovec2007,Lima-Mendez2009} would provide a common basis for understanding network assembly. And, such mechanisms could be used to create realistic synthetic networks for numerical simulations and experiments. Moreover, many studies have investigated how scale-free structure shapes the dynamics of processes running over the network. For example, scale-free random graphs are almost always susceptible to epidemics~\cite{Newman2002}, a result with profound implications for social influence and public health applications. The universality of scale-free networks is thus a critically important question, and resolving its empirical status would inform efforts to apply results from network theory to many scientific questions.

If real-world networks are not universally or even typically scale-free, the status of a unifying theme in network science over nearly 20 years~\cite{Albert1999, Barabasi1999,Carlson1999, Newman2005, Mitzenmacher2003} would be significantly diminished. Such an outcome would require a careful reassessment of the large literature that has grown up around the idea. Even if scale-free networks are not universal, their prevalence may be non-uniform across different domains of networks, e.g., it may be common for social networks but rare for biological networks to be scale free. Thus, a crucial followup question would be to assess this differential evidence by domain, 
and to characterize how real-world structures deviate from the scale-free pattern. For domains where there is little empirical support or where scale-free networks are relatively unusual, new models of structure and new mechanisms of assembly may be needed.

The validity and scope of the scale-free hypothesis is not uncontroversial. Some research has argued against its universality, on either statistical or theoretical grounds~\cite{Tanaka2005,Li2006, Przulj2007, Lima-Mendez2009,Willinger2009, Stumpf2012, Golosovsky2017}. There have also been passionate and personal defenses of its status~\cite{BarabasiIntro}, along with many findings claiming to validate or support its universality~\cite{Mislove2007, Agler2016, Goh2002, Pastor-Satorras2015, Gamermann2017}. This controversy has persisted because studies have typically relied upon small, often domain-specific data sets, less rigorous statistical methods, and differing definitions ``scale free"~\cite{Redner1998, Mislove2007, Agler2016, Goh2002,Ichinose2017, Pastor-Satorras2015, Zhang2015, Pachon2017}. Additionally, relatively few studies have performed statistically rigorous comparisons of a fitted power-law distribution to alternative, non-scale-free distributions, e.g., the log-normal or the stretched exponential, which can imitate a power-law form in realistic sample sizes~\cite{Clauset2009}.

For example, Eikmeier and Gleick~\cite{Eikmeier2017} recently investigated a number of real-world networks using rigorous methods for fitting and testing for the plausibility of power laws~\cite{Clauset2009} in the singular values of the adjacency matrix, the eigenvalues of the Laplacian, and the degree distribution. Although they argue that they find broad evidence of scale-free structure in these networks, the evidence remains ambiguous in two crucial ways. First, the statistical plausibility of scale-free structure is strongest in the singular and eigenvalue analyses rather than in the degree distributions, which is a different kind of scale-free structure than the hypothesis typically posits. Second, their analyses did not include controls for spurious conclusions due to small sample sizes or comparisons against alternative distributions, both of which are necessary to reduce the likelihood of false findings of scale-free structure~\cite{Clauset2009}.

Beyond statistical concerns, the literature on scale-free networks also exhibits conflicting interpretations of the theoretical meaning of observing a power-law degree distribution. Many studies interpret this pattern as direct evidence that a network was assembled by the preferential attachment mechanism, or by one of its variations. As a result, the term ``scale-free network" can confusingly refer to either a network with a power-law degree distribution or a network built by preferential attachment. This ambiguity has not helped resolve the controversy over the universality of scale-free networks, as a number of alternative mechanisms also produce power-law degree distributions without relying on preferential attachment~\cite{Carlson1999, Newman2005, Mitzenmacher2003}. This ambiguity highlights a simple fact:\ degree distributions reflect modest constraints on overall network structure~\cite{Alderson2007} and are thus weak measures by which to identify generating mechanisms~\cite{Mitzenmacher2004}.

Here, we carry out a broad test of the universality of scale-free networks by applying state-of-the-art statistical methods to a large and diverse corpus of real-world networks. To account for the variability in how scale-free networks have been defined in the literature, we formalize a set of quantitative criteria that represent differing strengths and types of evidence for scale-free structure in a particular network. These sets of criteria allow us to extend the scale-free hypothesis beyond the domain of simple (unweighted, undirected, static) graphs and thereby conduct a more complete evaluation of its universality. 

For each of 927 network data sets drawn from all domains of science, 
we estimate the best-fitting power-law model, test its statistical plausibility, and then compare it via a likelihood ratio test to alternative non-scale-free distributions. We analyze these results collectively,  
and then consider how the evidence for scale-free structure varies across domains. We conclude with a forward-looking discussion of the empirical status of the scale-free hypothesis and offer suggestions for future research on the structure of networks.

\section{Network data sets}

A key component of our evaluation of the scale-free hypothesis is the construction of a structurally and scientifically diverse corpus of real-world networks. We drew 927 network data sets from the \textit{Index of Complex Networks} (ICON), a comprehensive online index of research-quality network data sets, spanning all fields of science \cite{Clauset2016}. This corpus includes networks from biological, informational, social, technological, and transportation domains, and which range in size from hundreds to millions of nodes. 

\begin{figure}[t!]
\centering
    \includegraphics[width=0.48\textwidth]{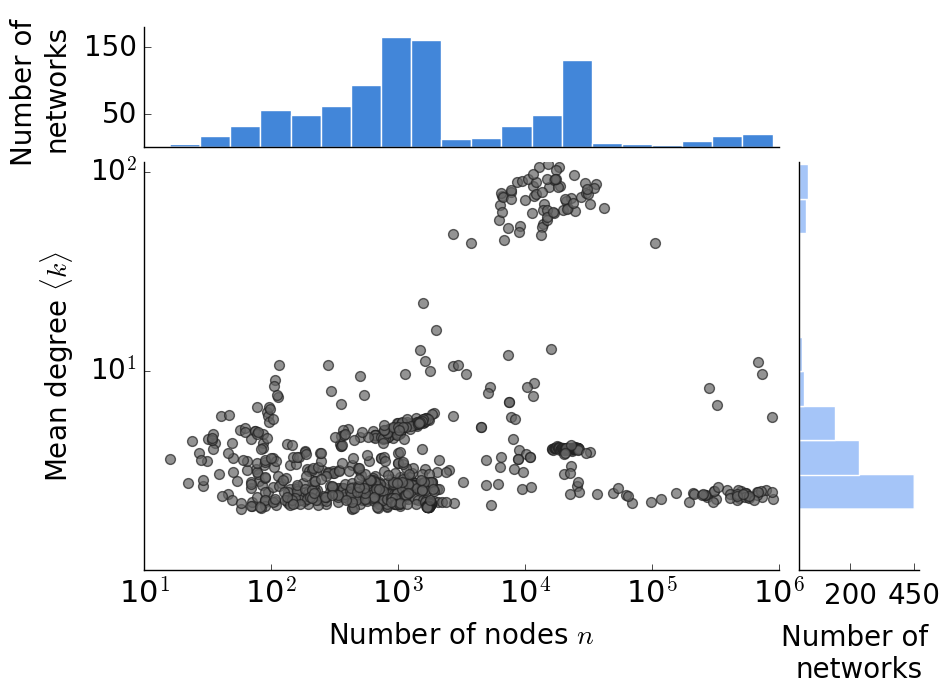}
  \caption{Mean degree $\langle k \rangle$ as a function of the number of nodes $n$ for the 927 network data sets studied here, showing a broad diversity of size and density. For data sets with more than one degree sequence (see text), we plot the median of the mean degrees of each corresponding degree sequence.}
    \label{degreedata}
\end{figure}

Because of natural class imbalance in the ICON index, the distribution of networks across domains within our corpus is not uniform (Table \ref{domaincounts}): roughly half are biological networks, a third are social or technological networks, and the remaining are informational or transportation networks. These networks span five orders of magnitude in size, are generally sparse, with a typical mean degree of $\langle k \rangle \approx 3$ (Fig.~\ref{degreedata}), and generally exhibit heavy-tailed degree distributions.

% ----- FIGURE 1 -----
% -------------------------

% ----- TABLE 1 -----
\begin{table}[t!]
\centering
\begin{tabular}{l | c | c c c c c c c }
Domain & Num(Prop) & Multip. & Bip. & Multig. & Weigh. & Dir. & Simp. \\ \hline
Bio. & 500 (0.54)  & 277 &  41& 384 & 29 & 37 &  39\\
Inf. & 15(0.02) &   0 &  0& 3&0& 5& 7\\
Social & 145(0.16) &  4   & 0& 5 & 3& 0&136 \\
Tech. & 200(0.21) & 122  & 0 & 3&1 & 192 &  5  \\
Trans. &    67 (0.070) & 48  & 0 & 65& 3&  2 & 0 \\  \hline
Total & 927(1.00) & 451 & 41 & 460 &36 & 236 & 187
\end{tabular}
\caption{Number of network data sets, and proportion of our network corpus, in each of five domains, under the taxonomy given by the \textit{Index of Complex Networks}~\protect\cite{Clauset2016}.}
\label{domaincounts}
\end{table}
% -------------------------

The scale-free property can be clearly defined for simple graphs, that is, for networks that are static, unipartite, unweighted, and undirected. For such networks, the degree distribution can be obtained unambiguously. However, many real-world networks are not simple, which complicates the extraction of a degree distribution.

\begin{figure}[t!]
\centering
    \includegraphics[width=0.48\textwidth]{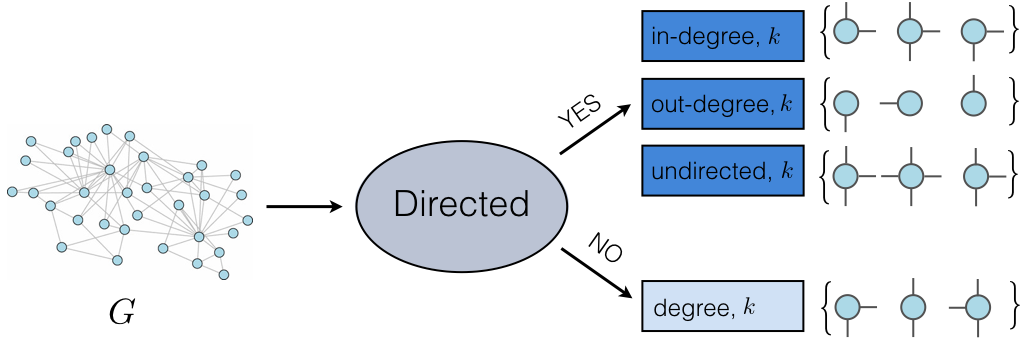}
  \caption{A graph simplification function, which takes as input a network $G$. In this case, if $G$ is directed, the function returns three degree sequences:\ the in-degrees, out-degrees, and undirected degrees, while if $G$ is undirected, it returns the degree sequence. Appendix~\ref{appendix:data} contains complete details.}
    \label{dataexample}
\end{figure}

Our ICON corpus contains many networks that have various combinations of directed, weighted, bipartite, multigraph, temporal, and multiplex properties. For each property, there can be multiple ways to extract a degree distribution and hence multiple ways to potentially test the scale-free hypothesis. We resolve these ambiguities by shifting perspectives and applying a consistent procedure. First, we define a \textit{network data set} to be a particular network in our corpus, e.g., a directed, weighted, multiplex network. Then, for each network data set, we extract a set of simple \textit{graphs} and their corresponding degree sequences, each of which can be tested for the scale-free property. Under this procedure, a directed network would produce three simple graphs and thus three degree sequences (Fig.~\ref{dataexample}), one for each of the in-degree sequence, the out-degree sequence, and the undirected degree sequence. Appendix~\ref{appendix:data} gives complete details of the procedure for extracting a set of simple graphs from a network data set with arbitrary properties.

% ----- FIGURE 2 -----
% -------------------------

Applying this procedure to our corpus of 927 network data sets, we obtain 4477 simple graphs that serve as the input to our statistical evaluation. We exclude from this set any graph with either very small mean degree $\langle k \rangle < 2$ or very large mean degree $\langle k \rangle > \sqrt{n}$. This filter removes 7376 extremely sparse and 12,146 extremely dense graphs, both of which are highly unlikely to be scale free. Some network data sets produce more simple graphs than others. For instance, an unweighted, directed temporal network with $T$ snapshots will produce $3(T + 1)$ simple graphs:\ 3 graphs for each of the $T$ snapshots, and 3 graphs for the time-aggregation of all the snapshots. Complicated network data sets can thus produce a combinatoric number of simple graphs. Hence, treating every graph independently in our evaluation of the scale-free hypothesis could lead to skewed results, e.g., if a few non-scale-free data sets account for a large fraction of the total extracted simple graphs. We control for this possibility by reporting our findings in terms of network data sets, rather than for individual graphs.

\section{Assessing Empirical Evidence}
All versions of the scale-free hypothesis include some kind of statement about a network's degree structure following a power-law distribution. Hence, a first methodological requirement for a broad evaluation of the empirical evidence for or against the universality of scale-free networks is the ability to test whether a particular degree sequence is consistent with a power-law distribution. For this question, we employ state-of-the-art statistical methods~\cite{Clauset2009}, which allow us to choose the best-fitting power law, test its statistical plausibility, and compare the quality of its fit to alternative non-scale-free distributions.

A second methodological requirement stems from our expansion of the scale-free hypothesis to non-simple networks. Because a given network data set can produce an entire set of degree sequences, we define five sets of criteria, which represent differing degrees of evidence that a particular network data set exhibits scale-free structure. Data sets that fail to meet any of these criteria are deemed Not Scale Free. Four sets of these criteria are nested, and represent the iterative addition of different characteristics, all found in the network science literature, about the properties of the power-law degree distributions. The fifth set represents the weakest possible version of the scale-free hypothesis, in which it is only required that the non-scale-free distributions are relatively worse models of the observed degrees than is the power law. Meeting this criterion does not provide direct evidence for the presence of scale-free structure, as networks that meet this criteria alone could still reject the power law as a direct model of their degree sequence.

\subsection{Fitting and comparing degree distribution models}
For each degree sequence $k_1,k_2,\dots,k_n$, we estimate the best-fitting power-law distribution model, where
\begin{align}
p(k) = C\, k^{-\alpha} \qquad \alpha > 1, \quad k\geq \kmin>0 \enspace,
\end{align}
meaning that the power-law form holds above some minimum value $k_{\rm min}$. In this form, $\alpha$ is the scaling exponent and $C$ is the normalization constant.

Degree distributions are integer-valued and hence we let $k$ be discrete. In the literature, it is commonly assumed that the power-law form may not hold for the smallest degree vertices. To allow for this possibility, we jointly estimate the parameter $\alpha$ for a power-law model of the upper tail of the degree distribution, and the minimum value $\kmin$ for which that model fits~\cite{Clauset2009}. 
Technical details of this estimation procedure are given in Appendix~\ref{appendix:PL}. 

Fitting the power-law distribution will always return some values for $\hat{k}_{\rm min}$ and $\hat{\alpha}$, but these give no indication of whether the fitted model is a statistically plausible explanation of the data. To assess this property of the fitted model, we use a goodness-of-fit test (see Appendix~\ref{appendix:PL}) to obtain a standard $p$-value. Following standard practice in this setting~\cite{Clauset2009}, if \mbox{$p\geq 0.1$}, then we deem the degree sequence to be plausibly scale free, while if \mbox{$p<0.1$}, we reject the scale-free hypothesis.

Once the power-law model has been fitted to the degree sequence, and its plausibility evaluated, we consider whether it is a better fit to the data than several alternative, non-scale-free distributions. This step is crucial because the inability to reject the power-law hypothesis does not guarantee that the power-law is the best description of a degree sequence. To determine whether a scale-free distribution is better than alternative heavy-tailed distributions, we compared the fitted power law to the (i) exponential, (ii) log-normal, (iii) power-law with exponential cutoff, and (iv) stretched exponential or Weibull distributions (Table \ref{lrt}), all of which have been used previously as models of degree distributions \cite{Amaral2000, Buzsaki2014, Jeong2001, Malevergne2005, DuBois2012}. Each alternative model is fitted via maximum likelihood to the empirical degrees $k\geq\hat{k}_{\rm min}$, with $\hat{k}_{\rm min}$ given by the power-law fit, which ensures that the comparison with the power-law model is not unfair~\cite{Clauset2009}.

The power-law fit is compared to alternatives using a likelihood ratio test (see Appendix~\ref{appendix:lrt}), which uses the difference in log-likelihoods between the power-law and alternative models as a test statistic:
\begin{align}
\mathcal{R} = \mathcal{L_{\text{PL}}} - \mathcal{L_{\text{Alt}}} \nonumber \enspace ,
\end{align}
where $\mathcal{L}_{\text{PL}}$ is the log-likelihood of the power-law model and $\mathcal{L}_{\text{Alt}}$ is the log-likelihood of a particular alternative model. The sign of $\mathcal{R}$ thus indicates which of the two models is favored, power law \mbox{($\mathcal{R}>0$)} or alternative \mbox{$(\mathcal{R}<0)$.} An important property of the likelihood ratio test, however, is a third outcome: $\mathcal{R}=0$, which indicates that the data do not permit a distinction between the models, i.e., neither is favored over the other.

Because $\mathcal{R}$ is derived from data, however, it is itself a random variable and thus subject to statistical variations \cite{Clauset2009, Vuong1989}. As a result, the sign of $\mathcal{R}$ is only meaningful if we can determine that $|\mathcal{R}|$ is far enough from $0$ to provide a clear conclusion. The standard solution for determining how to interpret $\mathcal{R}$ is another hypothesis test, in which we calculate a $p$-value against a null model of $\mathcal{R}=0$. If $p\geq0.1$, the sign of $\mathcal{R}$ is not informative and the data cannot tell us which model, power law or alternative, is a better fit. If $p<0.1$, then the data provide a clear conclusion in favor of one model or the other.

In order to report results at the level of an entire network data set, we apply the likelihood ratio tests to all the associated simple graphs and then aggregate the results. 
Specifically, for each alternative distribution, we count the number of simple graphs associated with a particular network data set in which the outcome favored the alternative, favored the power law, or had an inconclusive result.
Normalizing these counts across outcome categories provides a continuous measure of the relative evidence that the data set falls into each of category.

\subsection{Definitions of a scale-free network}
\label{section:sfdefs}
\begin{figure}[t!]
\centering
    \includegraphics[width=0.48\textwidth]{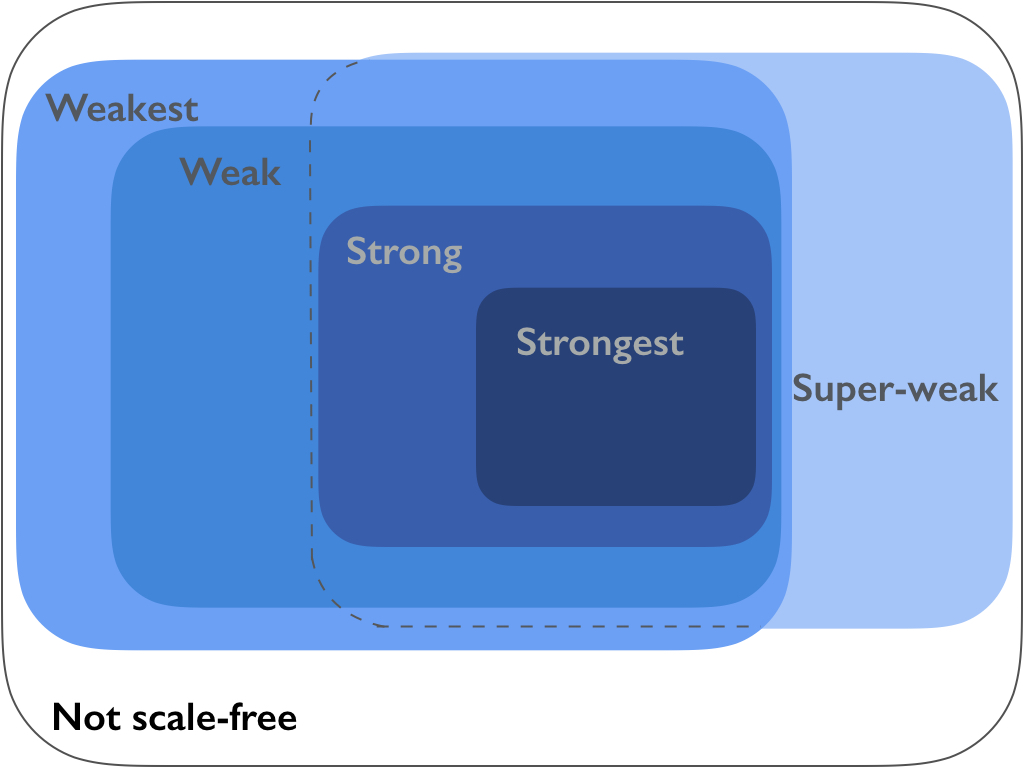}
  \caption{Taxonomy of scale-free network definitions, in order of increasing statistical evidence required:\ (i) \textit{Super-Weak} scale-free hypothesis, meaning that the power-law shape is not itself statistically plausible but is less implausible than alternatives; (ii) \textit{Weakest}, meaning plausibly power-law distributed; (iii) \textit{Weak}, adding a minimum of 50 observations in the tail; (iv) \textit{Strong}, also requiring $2< \hat{\alpha} <3$; and, (v) \textit{Strongest}, increasing the fraction of graphs that must fit the requirements. The \textit{Super-Weak} overlaps with the weak definitions and contains the strong definitions as special cases. Networks that fail to meet any of these criteria are deemed Not Scale Free. }
    \label{sfdefs}
\end{figure}

The methods described above can be applied to determine whether the degree sequence of a particular simple graph is convincingly scale free. However, because network data sets may produce many individual graphs as a result of simplification, the result for any one of several graphs is not sufficient evidence that the entire network data set is or is not convincingly scale free. Instead, the status of a network data set should depend on the results derived from all of its corresponding simple graphs. To capture the different levels and types of evidence of scale-free structure that our battery of tests may produce, we define five sets of criteria for a network data set (Fig.~\ref{sfdefs}).

The weakest-possible definition of a scale-free network, which we call \textit{Super-Weak}, is one in which the power-law form is merely a better description of the degree distribution than alternatives, regardless of whether the power law itself is a convincing model of the degrees. A network data set classified as Super-Weak is not necessarily scale free at all, and membership in this category simply indicates that the degree structure of the network data set is not obviously not scale-free. Formally, we define this category representing indirect evidence of scale-free structure as follows.
\begin{itemize}
\item {\bf{Super-Weak:}}
For at least $50\%$ of graphs, none of the alternative distributions are favored over the power law.
\end{itemize}
The four remaining definitions are nested, and represent increasing levels of direct evidence that the degree structure of the network data set is scale free:
\begin{itemize}
\item{\bf{Weakest:}}
For at least $50\%$ of graphs, the power-law hypothesis cannot be rejected ($p\geq 0.1$).
\item {\bf{Weak:}}
The requirements of the \textit{Weakest} set, and there are at least $50$ nodes in the distribution's tail ($\ntail>50$). 
\item {\bf{Strong:}} The requirements of the \textit{Weak} set, and that both $2< \hat{\alpha} < 3$, and for at least $50\%$ of graphs none of the alternative distributions are favored over the power-law. 
\item {\bf{Strongest:}} The requirements of the \textit{Strong} set for at least $90\%$ of graphs, rather than 50\%, and for at least $95\%$ of graphs none of the alternative distributions are favored over the power-law. 
\end{itemize}

% ----- FIGURE 3 -----
% -------------------------

The progression from Weakest to Strongest categories represents the addition of more specific properties of the power-law degree distribution, all found in the literature on scale-free networks or distributions. Broadly, these properties span requiring that a sufficient number of nodes be in the scaling regime, that the scaling parameter $\alpha$ fall in a specific range, and that the power-law model be favored over alternatives. Requiring a minimum of 50 points in the tail of the distribution is a fairly weak requirement to rule out cases where we are obviously fitting to noise.

In the Strongest category, the threshold of $90\%$ of graphs is determined by our choice of the Type I error rate for the goodness-of-fit hypothesis test. If all of the graphs for a network data set are indeed scale free, the goodness-of-fit test will be expected to incorrectly reject the power-law model $p=0.1$ of the time. The choice of $95\%$ for the comparison with the alternative distributions follows a similar rationale.

\subsection{Method validation on synthetic networks}
In order to evaluate the accuracy of these methods, we tested them on three classes of synthetic networks with mathematically well-understood degree distributions. Two mechanisms produce scale-free networks (a directed version of preferential attachment \cite{Easley2010} and a directed vertex copy model \cite{Newman2010}), and one does not (simple Erd\H{o}s-R\'enyi random graphs). Applied to synthetic networks generated by these mechanisms, our methodology correctly placed the synthetic data sets into scale-free categories suitable for their generating parameters, i.e., preferential attachment and vertex-copy data sets were categorized as scale-free networks, while Erd\H{o}s-R\'enyi random graphs were not (see Appendix \ref{appendix:synth}).

\section{Results}
We begin investigating the empirical evidence for the universality of scale-free networks by considering the distribution of estimated power-law scaling parameters $\Pr(\hat{\alpha})$, both for the corpus overall and for each group of data sets ranging from the Super-Weak to Strongest evidence categories. We then consider how the fitted power-law distributions fare relative to alternative distributions across data sets. Finally, we combine these results to form a quantitative assessment of the relative degrees of evidence for scale-free structure across all data sets in the corpus and by data sets drawn from different domains.

\subsection{Power-law distributions}

\begin{figure}[t!]
\centering
    \includegraphics[width=0.45\textwidth]{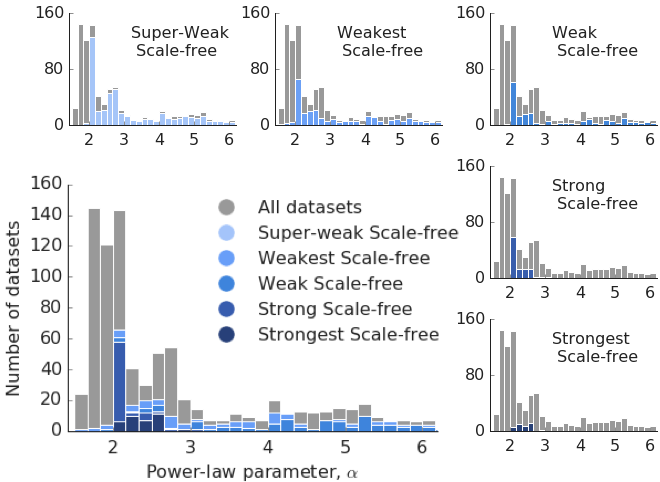}
  \caption{Distribution of median $\hat{\alpha}$-values by scale-free evidence category. For visual clarity $8\%$ of data sets, which had median $\hat{\alpha} \ge 6.5$, are omitted.}
   \label{alphadist}
\end{figure}

A simple summary of each network data set is the median scaling parameter $\hat{\alpha}$ among its corresponding simple graphs. Across the corpus, the overall distribution of median parameters is concentrated around a value of $\alpha=2$, but also exhibits a long right-tail, with $32\%$ of data sets having $\hat{\alpha}\geq 3$ (Fig.~\ref{alphadist}). The range $2<\alpha<4$ is sometimes identified as including the most emblematic of scale-free networks~\cite{Barabasi1999}, and we find that $43\%$ of network data sets have estimated parameters in this range. However, we also find that nearly $31\%$ of network data sets exhibit a median parameter in the range $\hat{\alpha}<2$, which is a relatively unusual value in the scale-free network literature.

% ----- FIGURE 4 -----
% -------------------------

\begin{figure}[t!]
  \centering
    \includegraphics[width=0.5\textwidth]{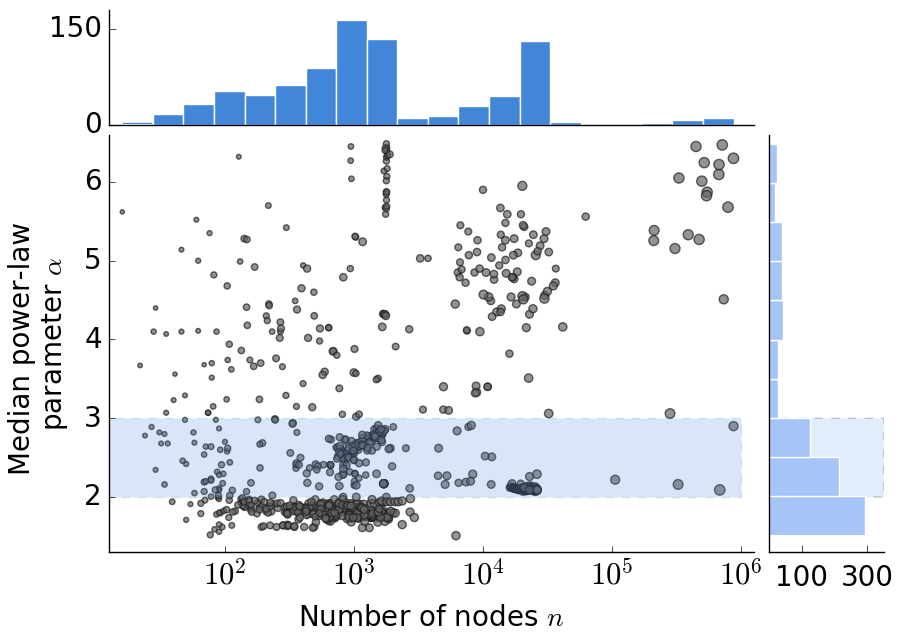}
    \caption{Median $\hat{\alpha}$ parameter as a function of network data set size $n$. A horizontal band highlights the canonical \mbox{$2 \leq \alpha \leq 3$} range and illustrates the broad diversity of estimated power-law parameters across empirical networks.}
     \label{scatter}
\end{figure}

However, the power-law distributions represented by these small parameters are not necessarily statistically plausible. Hence, the shape or concentration of this overall distribution is not evidence for, or against, the universality of scale-free networks. A simple check of our results is whether there is a clear relationship between the size of a network $n$ and the median power-law parameter $\hat{\alpha}$---a strong correlation, in either positive or negative directions, may be indicative of a systematic bias in our methodology. Instead, we find barely any correlation between $n$ and $\hat{\alpha}$, with $r^2=0.06$ ($p=4\times10^{-13}$), indicating little evidence of systematic bias (Fig.~\ref{scatter}).

% ----- FIGURE 5 -----
% -------------------------

Across the five categories of evidence for scale-free structure, the distribution of median $\hat{\alpha}$ parameters varies considerably (Fig.~\ref{alphadist}, insets). For data sets that fall into the Super-Weak category, we find a distribution with breadth similar to that of the overall distribution, with a long right-tail and many data sets with $\hat{\alpha} >3$. Only a handful of data sets with $\hat{\alpha} <2$ exhibit even Super-Weak or Weakest evidence of scale-free structure. In our corpus, these particular data sets correspond to planar networks, representing mycelial fungal or slime mold growth patterns~\cite{Lee2017}. Even among the Weakest and Weak categories, the distribution of median $\hat{\alpha}$ parameters remains broad, with a a substantial fraction exhibiting $\hat{\alpha}>3$. The Strong and Strongest categories require that $\hat{\alpha} \in [2,3]$, and we find that the few network data sets in these categories are slightly more prevalent near $\hat{\alpha}=2$.

\subsection{Alternative Distributions}
\label{section:altdists}
Independent of whether the power-law model is a statistically plausible explanation of a particular degree sequence, a model comparison between the fitted power law and alternative distributions can be instructive. In particular, the likelihood ratio test can reveal whether some other distribution is an equally good or even a better fit to the data. And, an outcome in which a power law is both a statistically plausible model of the data and a better model than alternatives provides strong evidence of scale-free structure. 

Across our corpus of network data sets, likelihood ratio tests find only modest support for the power-law distribution over alternatives (Table~\ref{lrt}). For instance, the exponential distribution, which exhibits a thin tail and relatively low variance, is favored over the power law (36\%) nearly as often as the power law is favored over the exponential (37\%). This outcome accords with the broad distribution of scaling parameters, as when $\alpha>3$ (Fig.~\ref{alphadist}, 32\% of data sets), the degree distribution must have a relatively thin tail.

The log-normal is a broad distribution that can exhibit very heavy tails, but which is nevertheless not scale free. Across data sets, we find that the log-normal is favored more than three times as often (45\%) over the power law, than vice versa (12\%), and the tests are inconclusive in the remainder of cases (43\%). Thus, in 88\% of cases, the log-normal was at least as good-fitting as the power law for degree distributions. This behavior reflects the well-known fact that log-normals and power laws are generally very difficult to distinguish with finite-sized samples~\cite{Clauset2009}.

The Weibull or stretched exponential distribution can produce both thin or heavy tails, and is a generalization of the exponential distribution. Like the log-normal, we find that the Weibull is more often a better statistical model of degree distributions (42\%) than is the power law (33\%). Finally, the power-law distribution with an exponential cut-off requires special consideration, as it contains the pure power-law model as a special case. As a result, the likelihood of the power law can never exceed that of the cutoff model, and the interesting outcome is the degree to which the test is inconclusive between the two. We find that just over half of the data sets favor the cutoff model over the pure power-law model, which suggests that, to the degree that scale-free networks are universal, finite-size effects in the extreme upper tail are quite common.

% ----- TABLE 2 -----
\begin{table}[t!]
\centering
\begin{tabular}{ l | c |ccc}
\multirow{2}{*}{\begin{tabular}[c]{@{}l@{}} \\ Alternative\end{tabular}} &\multicolumn{1}{c|}{\multirow{2}{*}{\begin{tabular}[c]{@{}l@{}} \\ $p(x)\propto f(x)$\end{tabular}} } & \multicolumn{3}{c}{Test Outcome} \\
 & \multicolumn{1}{c|}{}                                                                                     & $M_{\rm PL}$            & \quad  Inconclusive      \quad \quad        &  $M_{\rm Alt}$           \\ \hline
Exponential &       $\ds \textrm{e}^{-\lambda x}$     &  \bf{37\%}  &    27\%     &      36\%           \\
Log-normal & $\ds \frac{1}{ x}\textrm{e}^{{-\frac{\left(\log x-\mu \right)^2}{2\sigma^2}}}$   &   \bf{12\%}    &       43\%   &   45\%               \\
Weibull   & $\ds \textrm{e}^{-\left(\frac{x}{b}\right)^a}$ &   \bf{33\%}    &   25\%       &   42\%  \\
\multirow{2}{*}{\begin{tabular}[c]{@{}l@{}} Power law\vspace{-1mm} \\ with cutoff\end{tabular}} &  \multirow{2}{*}{\begin{tabular}[c]{@{}l@{}}  $\ds x^{-\alpha} \textrm{e}^{-\lambda x}$ \\ \end{tabular}}  &    \bf{---}   &    49\%     &  51\%       \\ \\
\end{tabular}
\caption{ Likelihood-ratio test results from comparing the best fit for four alternative distributions with the best fit power-law distribution for our data sets. We give the form of the distribution $f(x)$, and show the percentage of network data sets that favor the power-law model $M_{\rm PL}$, alternative model $M_{\rm Alt}$, or neither.  }
\label{lrt}
\end{table}
% -------------------------

\subsection{Assessing the Scale-free Hypothesis}
Given the results of fitting and testing the power-law distribution across network data sets, and the results of comparing that model to alternative models, we can now classify each according to the degree of evidence it exhibits for scale-free structure. Accordingly, each network data set is assigned to one of the five categories described in Section~\ref{section:sfdefs}, or classified as Not Scale Free.

\begin{figure}[t!]
  \centering
    \includegraphics[width=0.49\textwidth]{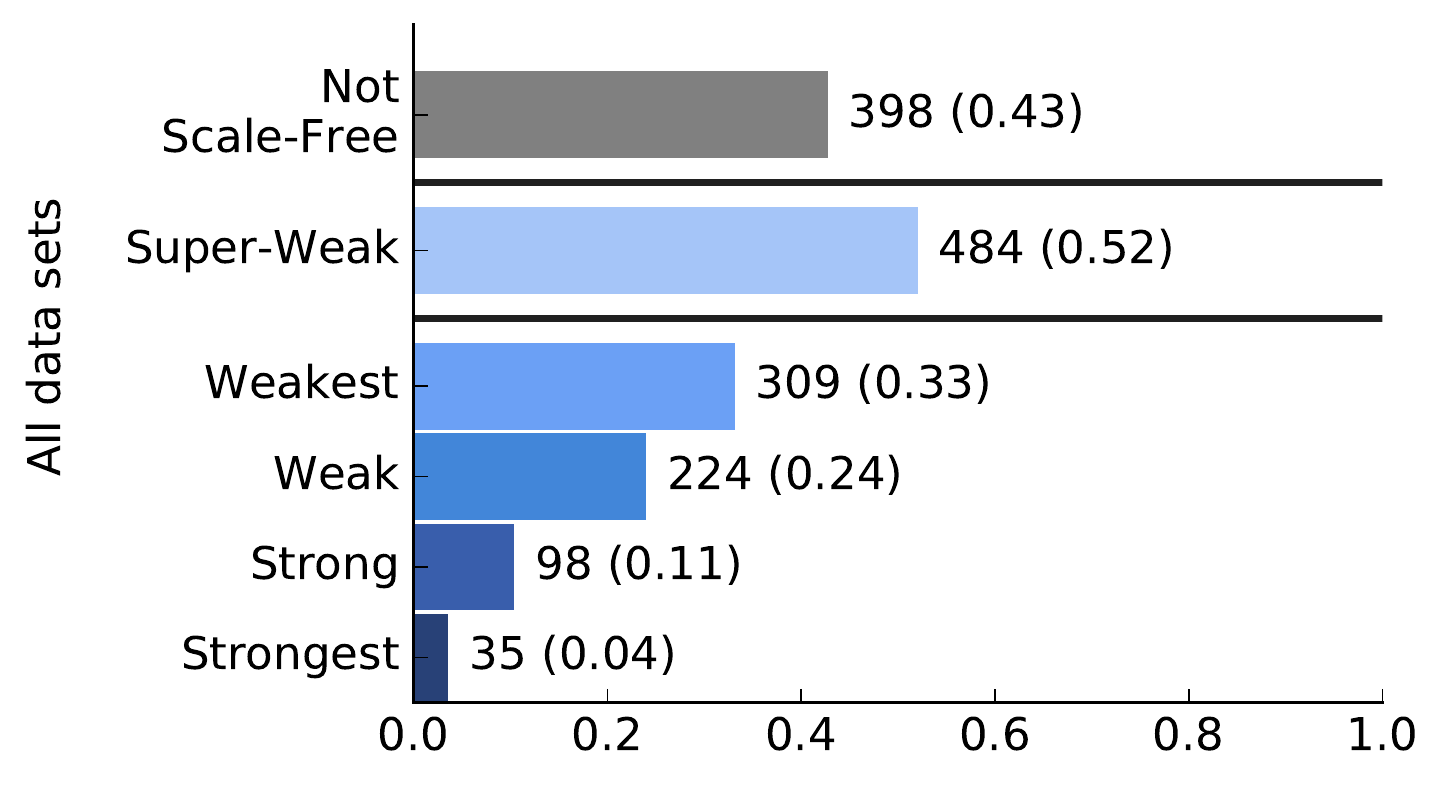}
  \caption{Proportions of networks in each scale-free evidence category. The four nested definitions are at the bottom, with super-weak separated by bars. Networks that are neither weakest or super-weak are considered not scale-free.}
    \label{sfall}
\end{figure}

Across our corpus, fully 43\% of network data sets fall into this latter category, admitting no evidence at all for scale-free structure (Fig.~\ref{sfall}). Slightly more than half (52\%) fall into the Super-Weak category, in which a scale-free pattern among the degrees is not itself statistically plausible, but remains marginally more plausible than alternative distributions. The Weakest and Weak categories represent network data sets in which the power-law distribution itself is a statistically plausible model of the networks' degree distributions. In the Weak case, this power-law scaling covers at least 50 nodes, a relatively modest requirement. These two categories account for 33\% and 24\% of data sets, respectively, indicating that direct statistical evidence of a plausibly scale-free degree distribution is, in fact, relatively uncommon.

Finally, only 11\% and 4\% of network data sets can be classified into the Strong or Strongest categories, respectively, in which the power-law distribution is not only a statistically plausible model of the degree structure, but the exponent falls within the special $\alpha\in[2,3]$ range and the power law is a better model of the degrees than alternatives. Taken together, these results indicate that genuinely scale-free networks are remarkably rare, and scale-free structure is not a universal pattern.

% ----- FIGURE 6 -----
% -------------------------

\begin{figure}[t!]
  \centering
    \includegraphics[width=0.49\textwidth]{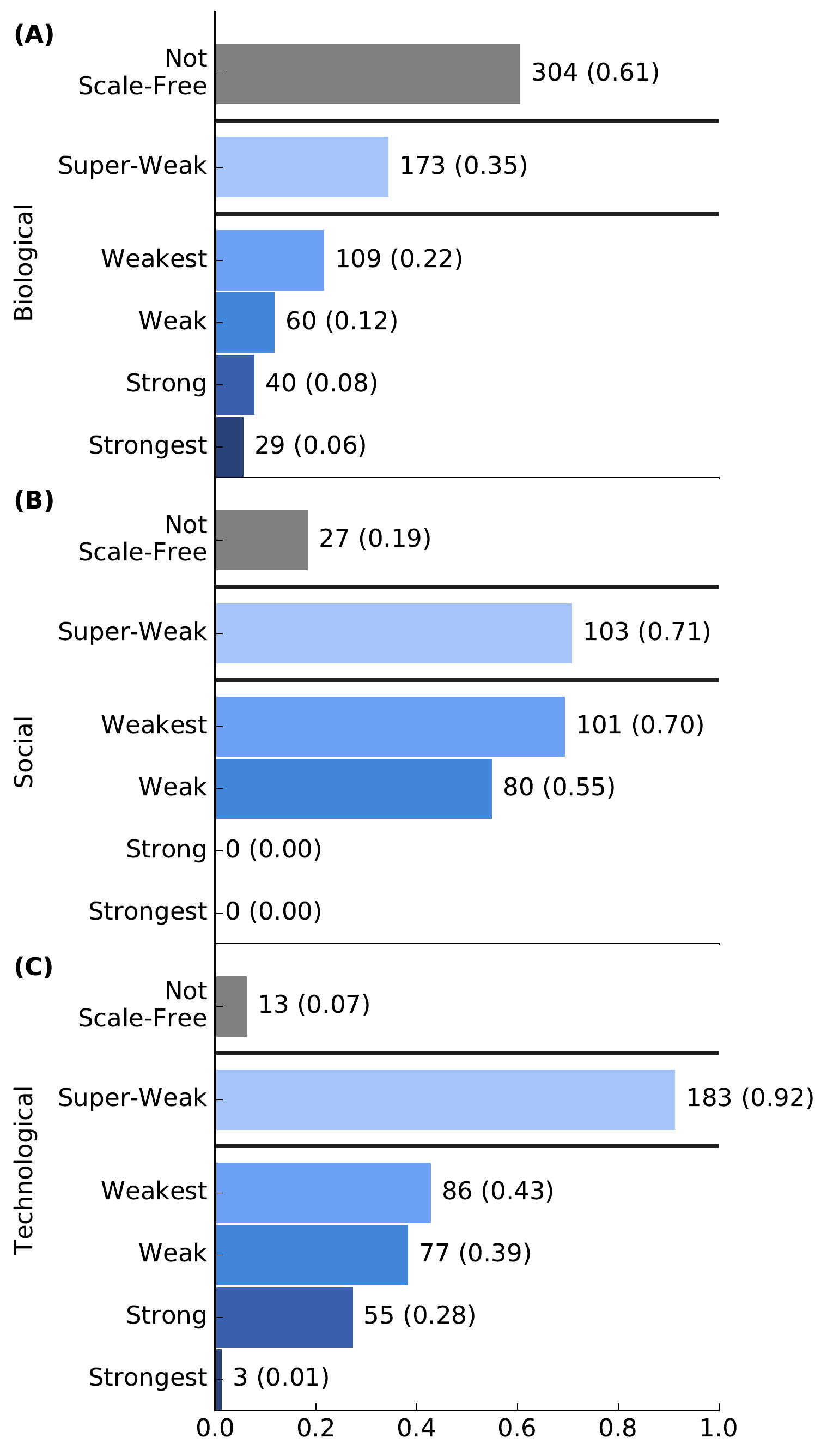}
  \caption{Proportions of networks in each scale-free evidence category divided by domain. (A) Biological networks. (B) Social networks. (C) Technological networks.}
    \label{sfdomains}
\end{figure}

The balance of evidence for or against scale-free structure in our corpus also varies by network domain (Fig.~\ref{sfdomains}). The degree to which different domains exhibit evidence of scale-free structure should clarify the past and shape future efforts to develop new structural mechanisms. Here, we focus our analysis on network data sets from three domains:\ biological, social, and technological, which represent 91\% of our corpus.

Among biological networks, a large majority lack any direct or indirect evidence of scale-free structure (61\% Not Scale Free; Fig.~\ref{sfdomains}a). The aforementioned fungal networks represent a large share of these Not Scale Free networks, but this group also includes some protein interaction networks and some food webs. Among the remaining networks, roughly one third exhibit only indirect evidence (35\% Super-Weak), and a modest fraction exhibit the weakest form of direct evidence (22\% Weakest). This latter group includes cat and rat brain connectomes. Compared to the corpus as a whole, biological networks are slightly more likely to exhibit the strongest level of direct evidence of scale-free structure (6\% Strongest), and these are primarily metabolic networks.

In contrast, social networks present a different picture, in two specific ways. First, a minority of social networks lack any direct or indirect evidence of scale-free structure (19\% Not Scale Free; Fig.~\ref{sfdomains}b), and a large majority exhibit indirect evidence (71\% Super-Weak). The former group includes the 2006 snapshot of collaborations in network science, and a number of both Facebook online social networks, 
and board of directors networks.
The Super-Weak group includes a large number of both Facebook friendship networks and board of director networks. 

And second, among the categories representing direct evidence of scale-free structure, not a single network data set falls into the Strong or Strongest categories. Hence social networks are at best only weakly scale free, with 70\% and 55\% exhibiting the weakest or weak direct evidence of scale-free structure, respectively. The social networks exhibiting weak evidence include about a third of the Facebook online social networks and half of the board of director networks.

Technological networks exhibit the smallest share of network data sets for which there is no evidence, direct or indirect, of scale-free structure (7\% Not Scale Free; Fig.~\ref{sfdomains}c), and the largest share exhibiting indirect evidence (92\% Super-Weak). The former group includes some digital circuit networks and water distribution networks. Among the categories representing direct evidence, less than half exhibit the weakest form of direct evidence (43\% Weakest). This group includes about half of the CAIDA autonomous systems networks, several peer-to-peer networks, and a few digital circuit networks.
In contrast to biological or social networks, however, technological networks exhibit a modest fraction of networks with strong direct evidence of scale-free structure (28\% Strong). Networks in this category include the other half of the CAIDA graphs. But, hardly any of the technological networks exhibit the strongest level of direct evidence (1\% Strongest).

% ----- FIGURE 7 -----
% -------------------------

\begin{figure}[t!]
  \centering
    \includegraphics[width=0.49\textwidth]{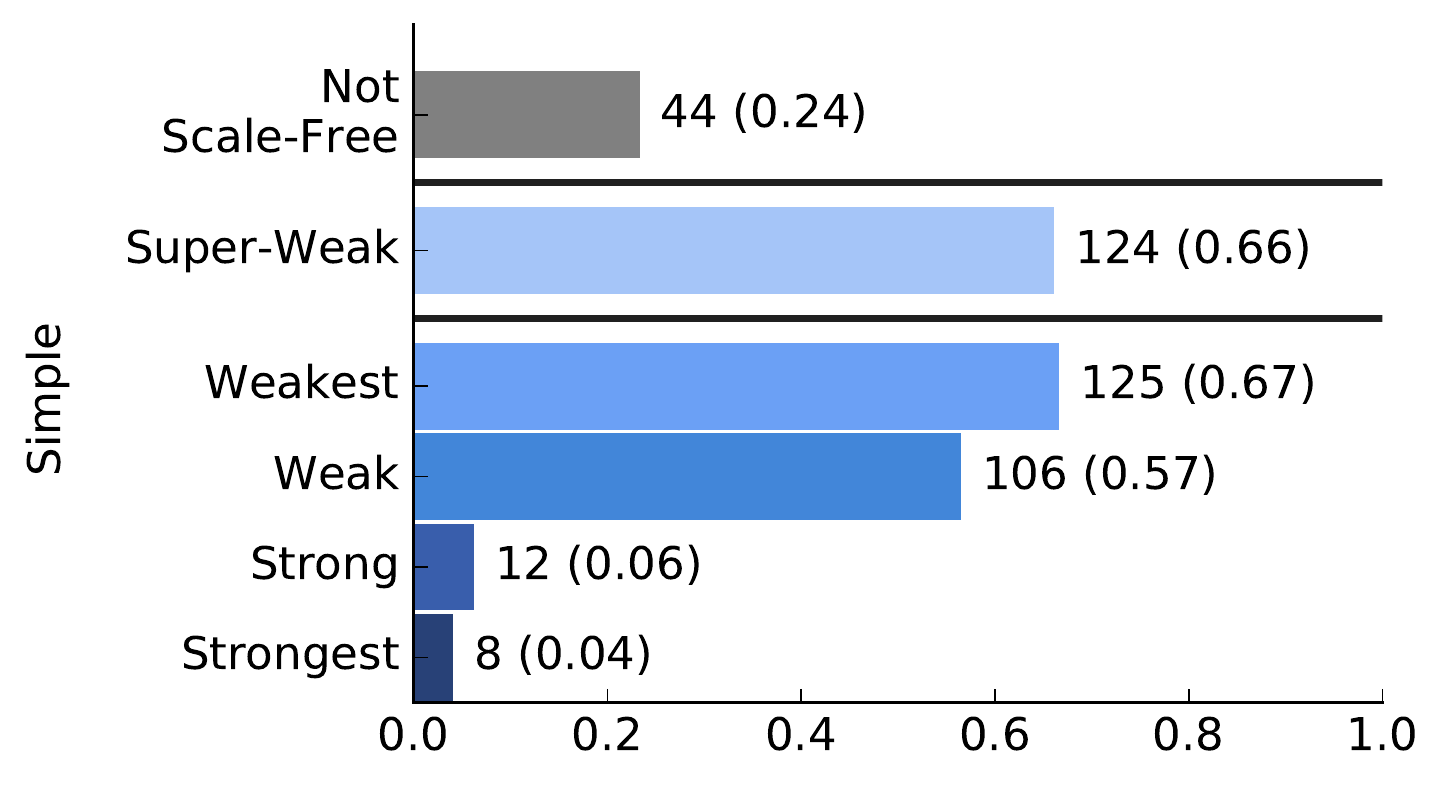}
  \caption{Proportions of networks in each scale-free evidence category for simple networks.}
    \label{sfsimple}
\end{figure}

Extending the scale-free network hypothesis to cover networks that were not simple allowed us to draw on a much larger range of empirical network data sets. It is therefore possible that the numerous non-simple network data sets we analyzed have distinct structural patterns from simple networks, and hence are less likely to exhibit a scale-free pattern. We test for this possibility by examining the classifications of the 187 simple networks within the corpus. Among these networks, a minority exhibit neither direct nor indirect evidence of scale-free structure (24\% Not Scale Free), and a modest majority exhibit at least indirect evidence (66\% Super-Weak; Fig.~\ref{sfsimple}). Compared to the overall corpus, about twice as many simple networks exhibit weak direct evidence, and about the same proportion exhibit strong direct evidence. These differences can be partly explained by the distribution of simple graphs by domain, as 73\% of simple graphs in the corpus are social, which exhibits similar proportions across the evidence categories. Hence, the general rarity of scale-free networks holds even when we restrict our analysis to simple graphs, and our inclusion of non-simple graphs has not skewed our results.

To rule out potential bias of our stronger scale-free definitions against the hypothesis, we also examine the results when we remove the power law with exponential cutoff from our list of alternative distributions. As the power law is a special case of the power law with cutoff, our likelihood-ratio test can only be inconclusive or result in favor of the power law with cutoff. In the case where the power law with cutoff is the best model, this case cannot be placed in the Strongest or Strong scale-free categories by definition. We found initially that 10.6\% of data sets fall into the Strong category. When we include data sets for which the power law with exponential cutoff was favored over the power law, this increases negligibly to 11.4\% of data sets. Additionally, if we also remove the restriction on the range of $\alpha$, the percentage of data sets in this Strong category increases to 23\%. This is very close to the results for the Weak category (24\%), which indicates that the majority of the decrease from the Weak to the Strong is due to the imposition of the bounds on $\alpha$ rather than the requirement against favoring alternative distributions. There is a similarly negligible increase in the number of datasets in the Strongest category, from 3.8\% to 4.4\% when we allow data sets for which the power law with exponential cutoff is favored. This is all consistent with the fact that the construction of our likelihood-ratio tests favors the power-law distribution since we use the $k_{\text{min}}$ that maximizes the likelihood of the power-law fit.

To summarize:\ across a large and diverse corpus, we find that it is remarkably rare for a network data set to exhibit the strongest form of direct evidence of scale-free structure, and this fact holds true across different domains. Recall that the Strongest level of evidence requires that the best-fitting power-law distribution (i) is itself statistically plausible and have an estimated parameter in the range $\alpha\in[2,3]$, and (ii) is at least as good a model as any alternatives. Hence the rarity of networks that meet these criteria implies that in general a scale-free distribution is rarely the best model of a network's degrees. Or, in other words, we find essentially no empirical evidence to support the special status that the power law has held in network science as a starting point for modeling and analyzing the structure of real networks. Instead, it is an empirical fact that real-world networks exhibit a rich variety of degree structures, relatively few of which are convincingly scale free.

% ----- FIGURE 8 -----

\section{Conclusions}

By evaluating nearly 1000 distinct real-world network data sets drawn from a wide range of scientific domains, we find that genuine scale-free networks are rare. In fact, less than 45 network data sets (4\%) exhibited the strongest level of direct evidence for scale-free structure, and only 33\% of network data sets exhibited the weakest form of direct evidence, in which a power law is at least a statistically plausible model of some portion of the upper tail of the degree distribution. Relaxing the criteria even further, to allow merely indirect evidence of scale-free structure in the degree distribution, admits only 52\% of networks data sets, and nearly half (43\%) of the data sets could be deemed Not Scale Free, meaning they showed no direct or indirect evidence of scale-free structure. These results clarify the empirical status of the common claim that scale-free networks are universal, and indicate that such claims are generally not empirically grounded.

Across different scientific domains, the evidence for scale-free structure is generally weak, but varies slightly in interesting ways. These differences provide some hints as to where genuine scale-free structure may exist. For instance, we find somewhat stronger evidence that scale-free structure occurs in some biological and technological networks. This pattern of evidence agrees with some theoretical work on domain-specific mechanisms for generating scale-free structure, e.g., in biological networks via the well-established duplication-mutation model for molecular networks~\cite{Pastor-Satorras2003, Ziv2005, Lima-Mendez2009} or in certain kinds of technological networks via highly optimized tolerance~\cite{Carlson1999,Newman:etal:2002}.

In contrast, we find that social networks are at best weakly scale free, indicating that if mechanisms like preferential attachment~\cite{Albert1999, Price1965, Simon1955} or its many variants~\cite{Leskovec2007, Berger2004} do operate in these contexts, they are strongly confounded by other mechanisms that cause statistically significant deviations from the expected power-law form. Class imbalance in the corpus prevents us from making broad conclusions about the prevalence of scale-free structure in informational networks or the theoretical relevance of scale-free mechanisms in that domain. However, the few informational networks we analyzed provided little indication that they would exhibit strongly different structural patterns that the domains with better representation in our corpus.

In general, the modest differences in the balance of evidence across the three domains for which we obtained sufficiently large samples---social, biological, and technological networks---seem to support a general conclusion that there is likely no single universal mechanism, or even a handful of mechanisms, that can explain the wide diversity of degree structures found in real-world networks. The failure to find universal patterns in the degree structure of networks indicates that a great deal remains unknown about how networks from different domains differ and what kind of structural patterns are common within them. We look forward to new investigations that detail these differences and commonalities.

The empirical rarity of scale-free networks presents both a puzzle and an opportunity. The strong focus on scale-free patterns over nearly 20 years has meant relatively less is known about alternative mechanisms that produce non-scale-free structural patterns. Hence, an important direction of future work in network theory will be development and validation of novel mechanisms for generating more realistic degree structure in networks. Similarly, theoretical results concerning the behavior of dynamical processes running on top of networks, including spreading processes like epidemiological models or influence models, may need to be reassessed in light of the genuine structural diversity of real-world networks.

The statistical methods and evidential categories developed and used in our investigation provide a quantitatively rigorous manner by which to assess whether some network exhibits scale-free structure or not. Their application to a novel network data set should enable future researchers to determine whether scale-free modeling assumptions are empirically justified.

Furthermore, large corpora of real-world networks, like the one used here, represent a powerful, data-driven resource by which to investigate the structural variability of real-world networks. Such corpora could be used to evaluate the empirical status of many other broad claims in the networks literature, including the tendency of social networks to exhibit high clustering coefficients and positive degree assortativity~\cite{Newman:Park:2003}, the prevalence of the small-world phenomena~\cite{Watts:Strogatz:1998}, the prevalence of ``rich clubs'' in networks~\cite{Colizza:etal:2006}, the ubiquity of community~\cite{Girvan:Newman:2002} or hierarchical structure~\cite{Clauset:etal:2008}, and the existence of ``super-families'' of networks~\cite{Milo:etal:2004}. We look forward to those investigations and the new insights they are sure to bring to our understanding of the structure and function of networks.

\begin{acknowledgements}
The authors thank Eric Kightley and Johan Ugander for helpful conversations, and acknowledge the BioFrontiers Computing Core at the University of Colorado Boulder for providing High Performance Computing resources (NIH 1S10OD012300) supported by BioFrontiers IT. This work was supported in part by Grant No. IIS-1452718 (A.C.) from the National Science Foundation. Code for graph simplification functions and power-law evaluations, as well as replication data, are available at \url{https://github.com/adbroido/SFAnalysis}.
\end{acknowledgements}

\section*{Author contributions}
A.D.B. and A.C. conceived the research, designed the analyses, and wrote the manuscript. A.D.B. conducted the analyses.

%---------------------BIBLIOGRAPHY-----------------------------------
\bibliographystyle{apsrev4-1}
\bibliography{test}
%-------------------------------------------------------------------------

\appendix
\section{Simplifying Network Data Sets}
\label{appendix:data}
\begin{figure}[t!]
  \centering
    \includegraphics[width=0.5\textwidth]{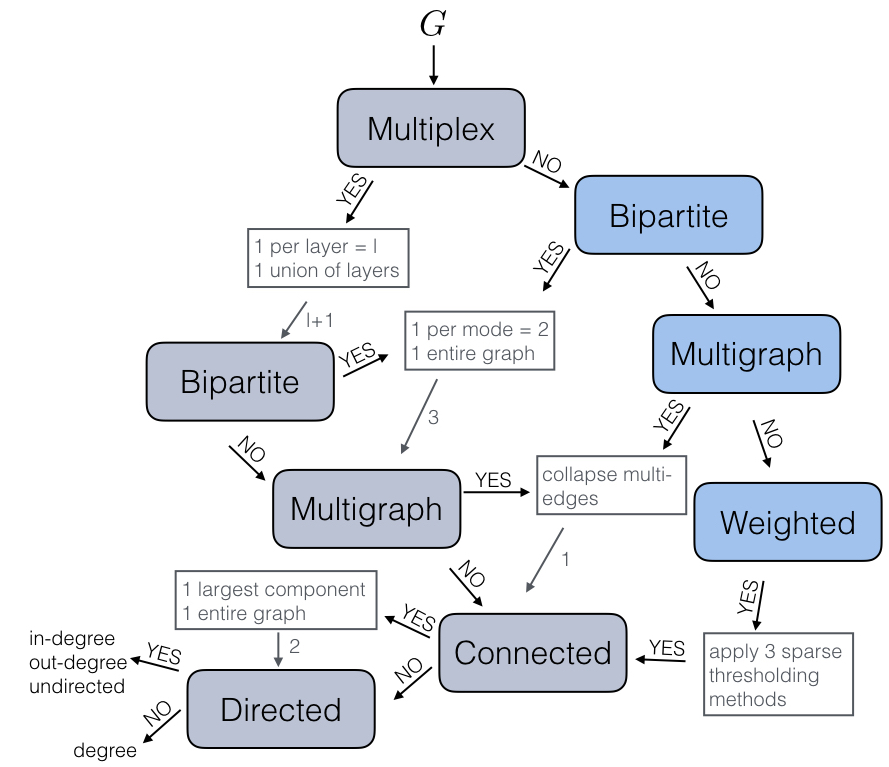}
  \caption{Flowchart describing the path from network data set to degree sequence(s). Each step removes a layer from the properties. The gray path is for multiplex, bipartite, or multigraph networks, while the blue is for weighted networks without these properties. Details in text.}
  \label{flowchart}
\end{figure}

Our corpus of real-world networks includes both simple graphs and networks with various combinations of directed, weighted, bipartite, multigraph, temporal, and multiplex properties. For each property, there can be multiple ways to extract a degree sequence, and in some cases, extracting a degree sequences requires making a choice. To resolve these ambiguities, and to provide a consistent set of rules by which to extract a set of degree sequences from a given network data set, we developed a set of graph simplification functions, which are applied in a sequence that depends on the graph properties of the input (Fig.~\ref{flowchart}). For completeness, we describe these specific pathways, and give counts of how many network data sets in our corpus followed each pathway.

% ----- FIGURE 9 -----
% -------------------------

At each stage in our processing we remove one property at a time, making the network simpler and never adding properties. Repeating this process for each property in succession converts a network data set into a set of simple graphs. Some networks are processed into a large number of simple graphs, due to combinatoric blowup of multiple properties. To minimize some of this, we treat weighted graphs differently depending on whether or not they have any multiplex, bipartite, or multigraph properties. Multiplex networks include temporal networks, so many of these have a large number of layers, which can generate many simple graphs, (see below). Thus to mitigate some of the potential blowup, if the graph has any of these properties we ignore the weights. If not, however, the data set is replaced with three unweighted graphs as follows. The goal of this transformation is to replace a potentially dense weighted graph with a set of unweighted graphs that are relatively sparse, but not so sparse as to be trivially disconnected so we identify and then apply three thresholds to the edge weights, so that the resulting unweighted graphs have $\langle k \rangle =\{2, n^{1/4},\sqrt{n}\}$. These threshold values are determined by the empirical edge weight distribution of the graph, and correspond to choosing the $m=\{n, (1/2)n^{5/4}, (1/2)n^{3/2}\}$ largest-weight edges, respectively. 

The lower value of $\langle k \rangle$ or $m$ represents a very sparse graph, retaining primarily the largest-weight edges, but not so sparse as to be likely strongly disconnected. The upper value represents a relatively dense network, retaining all but the smallest-weight edges, but not so dense that the degree distribution is trivial. The middle value splits the difference between these.
Our corpus contains only 3 weighted networks, 5 weighted directed networks, for 8 total weighted networks, so these networks represent a modest share of the resulting corpus.

Multiplex and temporal network data sets are composed of $T$ ``layers,'' each of which is a network itself. The multiplex network is replaced by a set of $T+1$ graphs, one for each layer and one for the union of edges and nodes across all layers. In this way, the multiplex or temporal property is removed, and the original data set replaced with a set of graphs. Each graph in this set is then further processed to remove any remaining non-simple properties. A bipartite graph is replaced with three graphs:\ one each for the ``A-mode'' projection, ``B-mode'' projection, and original bipartite graph. If present, multi-edges are collapsed and weights discarded. Next, a graph is checked to determine whether it is connected. If it is not connected, it is replaced with two graphs:\ one which contains only the largest component and one that contains the entire, disconnected graph.

As a final step, directed graphs are replaced by three degree sequences:\ one for the in-degrees, one for the out-degrees, and one for the total degrees; undirected graphs are replaced with the single degree sequence. The results of this sequential processing is a set of degree sequences that, as a group, represent the original network. Our corpus contains 2 pure multiplex networks, 320 multiplex multigraphs, and 130 multiplex directed networks; this yields 451 total multiplex networks.

Network data sets that are bipartite and not multiplex are first replaced with three graphs:\ one for the ``A-mode'' project, one for the ``B-mode'' projection, and one for the original bipartite graph. Each of these graphs is then processed starting from just after the bipartite step described above in the multiplex or temporal network processing pathway. In our corpus, there are 16 purely bipartite networks, and 25 bipartite weighted networks; this yields 41 bipartite networks total.

Data sets that are multigraphs, but not multiplex/temporal or bipartite, are merely simplified by collapsing multi-edges. Edge weights are then discarded, and the resulting graph is processed starting from the check for directedness as above. In our corpus, there are 137 multigraphs and 2 weighted multigraphs, 1 weighted directed multigraph; this yields 460 multigraphs total, including those that are multiplex.

Data sets that are only directed, with no other properties, are first checked for their connectedness (see above) and are then processed to produce three degree sequences:\ one each for the in-degrees, out-degrees, and total degrees. In our corpus, there are 236 directed networks. In the case of a simple graph, the only check is for connectedness. Our corpus contains 187 simple networks.

Replication data, in the form of the corpus of degree sequences obtained by the above simplification steps, is available (see acknowledgements). The corpus of 927 original network data sets represents approximately 250GB of data, and is hence not easily shareable.

\section{Power-law analysis}
\label{appendix:PL}
\subsection{Fitting the model}
If $k$ follows a discrete power-law distribution starting at $\kmin\ge 1$, then pdf of the power law has the form
\begin{align}
p(k) = \frac{1}{\zeta(\alpha, \kmin) } \sum_{k=\kmin}^{\infty} k^{-\alpha} \nonumber
\end{align}

where $\ds \zeta(\alpha, \kmin) = \sum_{i=0}^\infty (i+\kmin)^{-\alpha}$ is the Hurwitz zeta function.

Estimating $\alpha$ requires first choosing $\kmin$, which we estimate via the standard Kolmogorov-Smirnov (KS) minimization approach~\cite{Clauset2009}. This methods selects the $\kmin$ that minimizes the maximum difference in absolute value between the (cumulative) empirical distribution $E(k)$ on the observed degrees $k\geq \kmin$ and the cdf of the best fitting power law $P(k\,|\,\hat{\alpha})$ on those same observations. This difference, called the KS-statistic, is defined as
\begin{align}
D = \max_{k\ge \kmin} \left| E(k) - P(k\,|\,\hat{\alpha}) \right| \nonumber \enspace.
\end{align}
We choose as $\kmin$ the value that minimizes the $D$. The estimate $\hat{\alpha}$ is chosen by maximum likelihood (the MLE), which we obtain by numerically optimizing the log-likelihood function~\cite{Clauset2009}.

\subsection{Testing goodness-of-fit}
We assess the goodness-of-fit of the fitted model using a standard $p$-value, numerically estimated via the standard semi-parametric bootstrap approach~\cite{Clauset2009}.
Given a degree sequence with $n$ elements, of which $n_{\rm tail}$ are $k\ge \kmin$ and with MLE $\hat{\alpha}$, a synthetic data set is generated as follows. For each of $n$ synthetic values, with probability 
$n_{\rm tail}/n$ we draw a random number from the fitted power-law model, with parameters $\kmin$ and $\hat{\alpha}$. Otherwise, we 
choose a value uniformly at random from the empirical set of degrees $k<\kmin$. Repeated $n$ times this produces a synthetic data set that closely follows the empirical distribution below $\kmin$ and follows the fitted power-law model at and above $\kmin$.

Applying the previously defined fitting procedure to a large number of these synthetic data sets yields the null distribution of KS-statistics $\Pr(D)$. Let $D^{*}$ denote the value of the KS-statistic for the best fitting power-law model for the empirical degree sequence. The $p$-value for this model is defined as the probability of observing, under the null distribution, a KS-statistic at least as extreme as $D^{*}$. Hence, $p=\Pr(D\geq D^{*})$ is the fraction of synthetic datasets with KS statistic larger than that of the empirical data set.
Following standard practice, if $p<0.1$, then we reject the power law as a plausible model of the degree sequence, and if $p\ge0.1$, then we fail to reject the model~\cite{Clauset2009}. We note:\ failing to reject does not imply that the model is correct, only that it is a plausible data generating process.

\section{Alternative Distributions}
\label{appendix:alt}
\subsection{Exponential}
If $k$ follows a discrete exponential distribution starting at $\kmin$, then the pdf of the exponential has the form

\begin{align}
p(k) = \left(\frac{\textrm{e}^{-\lambda \kmin}}{1-\textrm{e}^{-\lambda}}\right) \textrm{e}^{-\lambda k} \nonumber  \enspace .
\end{align}
As with the power-law distribution, we use standard numerical maximization routines to estimate the maximum likelihood choice of $\lambda$.

\subsection{Log-normal}
The log-normal distribution is typically defined on a continuous variable $k$. To adapt this distribution to discrete values, we bin the continuous distribution and then adjust so that it begins at $\kmin$ rather than at $0$.

Let $f(k)$ and $F(k)$ be the density and distribution functions of a continuous log normal variable, where
\begin{align}
f(k) = \frac{1}{\sqrt{2\pi}\sigma k}\textrm{e}^{-\frac{\left(\log k-\mu \right)^2}{2\sigma^2}} \quad, \quad x>0 \nonumber
\end{align}
and
\begin{align}
F(k) = \frac{1}{2 }+\frac{1}{2 }\mbox{erf}\left[\frac{\left(\log k-\mu \right)}{\sqrt{2}\sigma}\right] \enspace . \nonumber
\end{align}
We define $g(k)$ and $G(k)$ to be the density and distribution functions of a discrete log-normal variable, given by
\begin{align}
g(k) = F(k+1)-F(k) \quad, \quad x\ge0 \nonumber
\end{align}
and
\begin{align}
G(k) = \sum_{y=0}^{k}g(y) = F(k+1)-F(0) = F(k+1) \enspace . \nonumber
\end{align}

We then generalize the distribution to start at some minimum value, i.e., rather than starting at 0, the distribution starts at $k=\kmin$, where $\kmin$ is a positive integer. This pmf is obtained by re-normalizing the tail of $g(k)$ so that it sums to 1 on the interval $\kmin$ to $\infty$, yielding
\begin{align}
h(k) &= \frac{g(k)}{\sum_{k=\kmin}^{\infty}g(k)}  = \frac{g(k)}{1-\sum_{k=0}^{\kmin-1}g(k)} \nonumber \\
& = \frac{g(k)}{1-G(\kmin-1)}  = \frac{g(k)}{1-F(\kmin)} \nonumber \enspace .
\end{align}

Maximum likelihood estimation was carried out using standard numerical optimization routines. 
Additionally, we constrained the optimization in order to prevent numerical instabilities. Specifically, we required $\sigma\geq 1$ and $\mu\geq -\lfloor n/5\rfloor$. As a check on these constraints, we verified that in no cases did the likelihood improve significantly by allowing $\sigma < 1$, and the constraint on $\mu$ prevents it from decreasing without bound (a behavior that can produce arbitrarily heavy-tailed distributions over a finite range in the upper tail).
To initialize the numerical search, we set $(\mu_0, \sigma_0) = (0,1)$.

\subsection{Power-law with exponential cutoff}
If $k$ follows a discrete power-law distribution starting at $\kmin$, and with an exponential cutoff in the upper tail, then its pdf has the form
\begin{align}
p(k) = \left[ \textrm{e}^{-\kmin\,\lambda}\,\Phi(\textrm{e}^{-\lambda},\alpha,\kmin) \right] k^{-\alpha} \textrm{e}^{-\lambda k} \nonumber 
\end{align}

where $\ds \Phi(z,s,a) =\sum_{i=0}^\infty \frac{z^i } {(a+i)^s }$ is the Lerch Phi function. 

We again estimate this distribution's parameters $\lambda$ and $\alpha$ using standard numerical maximization routines.

\subsection{Weibull (Stretched exponential)}
A common approach to obtain a discrete version of the stretched exponential or Weibull distribution is to bin the continuous distribution~\cite{Nakagawa1975}. Let $f(k)$ and $F(k)$ be the density and distribution functions of a continuous Weibull variable, where
\begin{align}
F(k) = 1-\textrm{e}^{-\left(k/b\right)^a}, \quad x\ge0 \nonumber \enspace .
\end{align}

Define $g(k)$ and $G(k)$ to be the density and distribution functions of a discrete Weibull variable, given by:
\begin{align}
g(k) = F(k+1)-F(k), \quad x\ge0 \nonumber
\end{align}
and
\begin{align}
G(k) = \sum_{y=0}^{k}g(y) = F(k+1)-F(0) = F(k+1) \enspace . \nonumber
\end{align}

As with the log-normal, we generalize the distribution to start at some minimum value, i.e., rather than starting at 0, the distribution starts at $k=\kmin$, where $\kmin$ is a positive integer. This pmf is obtained by re-normalizing the tail of $g(k)$ so that it sums to 1 on the interval $\kmin$ to $\infty$, yielding
\begin{align}
h(k) 
&= \textrm{e}^{\left(\kmin/b\right)^a}\left[\textrm{e}^{-\left(k/b\right)^a}-\textrm{e}^{-\left((k+1)/b\right)^a}  \right] \enspace . \nonumber
\end{align}
As with the other distributions, we estimate this distribution's parameters using standard numerical maximization routines.

\section{Likelihood-ratio tests}
\label{appendix:lrt}
The power-law models were compared with the alternatives using a set of likelihood-ratio tests. For each alternative distribution, we obtained the log-likelihood $\mathcal{L_{\text{Alt}}}$ of the best fit. The difference between this value and the log-likelihood of the power-law fit to the same observations yields the likelihood ratio test (LRT) statistic $\mathcal{R} = \mathcal{L_{\text{PL}}} - \mathcal{L_{\text{Alt}}}$.

When $\mathcal{R} > 0$, the power law is a better fit to the data. Similarly, when $\mathcal{R}<0$, the alternative distribution is the better fitting model. Crucially, when $\mathcal{R}=0$, the test is inconclusive, meaning that the data cannot distinguish between the two models. The test statistic $\mathcal{R}$, however, is itself a random variable, and hence is subject to statistical fluctuations. Accounting for these fluctuations dramatically improves the accuracy of the test by reducing both types of incorrect decision rates~\cite{Clauset2009}.
As a result, the sign of $\mathcal{R}$ alone is not a reliable indicator of which model is a better fit. The now standard approach for controlling for this uncertainty is to calculate a $p$-value against the null model of $\mathcal{R}=0$. Only if that model can be rejected, is the sign of $\mathcal{R}$ meaningful~\cite{Vuong1989}. In this setting, if $p<0.1$, then the absolute value of $\mathcal{R}$ is sufficiently far from 0 that its sign is interpretable.

We obtain this $p$-value with the same method used in Ref.~\cite{Clauset2009}, originally proved valid in Ref.~\cite{Vuong1989}. Note that 
\begin{align*}
\mathcal{R} &= \mathcal{L_{\text{PL}}} - \mathcal{L_{\text{Alt}}} \\
&= \sum_{i=1}^n\left[\ln{p_{\text{PL}}(k_i)}-\ln{p_{\text{Alt}}(k_i)}\right] \\
& =  \sum_{i=1}^n \left[\ell_i^{(\text{PL})}-\ell_i^{(\text{Alt})}\right]
\end{align*}
where $\ell_i^{(\text{PL})}$ is the log-likelihood of a single observed degree value $k_i$ under the power-law model, and $n$ is the number of empirical observations being used by a model (in our setting, this number is $n_{\rm tail}$, but we omit that annotation to keep the mathematics more compact).

We have assumed that the degree values $k_i$ are independent, which means the point-wise log-likelihood ratios $\ell_i^{(\text{PL})}-\ell_i^{(\text{Alt})}$ are independent as well. The central limit theorem states that the sum of independent random variables becomes approximately normally distributed as their number grows large, and that this normal distribution has mean $\mu$ 
and variance $n\sigma^2$, where $\sigma^2$ is the variance of a single term. This distribution can be used to obtain the $p$-value, but requires that we first estimate $\mu$ and $\sigma^2$. Note that we assume $\mu = 0$ because the null hypothesis is $\mathcal{R}=0$. 
We then approximate $\sigma^2$ as the sample variance in the observed $\mathcal{R}$
\begin{align}
\sigma^2 = \frac{1}{n-1} \sum_{i=1}^n \swrp{\wrp{\ell_i^{(\text{PL})}-\ell_i^{(\text{Alt})}} -\wrp{\bar{\ell}_i^{(\text{PL})}-\bar{\ell}_i^{(\text{Alt})}} }^2 \enspace , \nonumber
\end{align}
where 
\begin{align}
\bar{\ell}_i^{(\text{PL})} = \frac{1}{n}\sum_{i=1}^n l_i^{(\text{PL})} \quad \text{ and } \quad \bar{\ell}_i^{(\text{Alt})} = \frac{1}{n}\sum_{i=1}^n \ell_i^{(\text{Alt})} \nonumber
\end{align}
are sample means.

Under this null distribution, the probability of observing an absolute value of $\mathcal{R}$ at least as large as the actual test statistic is given by the two-tail probability
\begin{align}
p = \frac{1}{\sqrt{2\pi n\sigma^2}}\swrp{\int_{-\infty}^{-|\mathcal{R}|} \!\!\textrm{e}^{-\frac{t^2}{2n\sigma^2}}\, \textrm{d}t + \!\! \int_{|\mathcal{R}|}^{\infty} \!\! \textrm{e}^{-\frac{t^2}{2n\sigma^2}}\, \textrm{d}t}.
\end{align}

Hence, following standard practice~\cite{Clauset2009}, if $p \le 0.1$, then we reject the null hypothesis that $\mathcal{R}=0$, and proceed by interpreting the sign of $\mathcal{R}$ as evidence in favor of one or the other model.

\section{Evaluating the method on synthetic data with ground truth}
\label{appendix:synth}
We tested the method on synthetic data sets to see if we recover ground truth structure. 
We generated 100 random 5000-node networks by each of three methods.
We expect two of the methods, preferential attachment \cite{Easley2010} and vertex copying \cite{Newman2010}, to generate scale-free networks, and Erd\H{o}s-R\'enyi random graphs not to.

The first method, preferential attachment \cite{Easley2010}, is one of the most commonly referenced scale-free generating mechanisms. The process is as follows. We begin with a 4-node directed network, in which each node has 3 out edges, one to each of the other nodes. We then add one node at a time until we reach a total of 5000 nodes in the network. Each added node forms 3 out edges. For each edge, with probability $p=2/3$ the connection is formed preferentially. That is, the new node connects to an existing node with probability proportional to the in-degree of that node. With probability $1-p=1/3$, the connection is uniform, where each existing node has equal probability of receiving the new edge. We expect the in-degree sequence of the final graph to follow a power-law distribution, while the out degree of every node is 3. In our processing, we extract the in-, out-, and total-degree sequences. We'd expect that most of the time the total-degree sequence will also look power-law, since it is the sum of a power law and a constant distribution.

We find 87\% of the preferential attachment graphs fall into the Super-Weak category.  Further, if we do not consider the power law with cutoff as an alternative model, this increases to 98\%. This means 98\% of the time, a power-law model is favored over alternatives for these graphs. When we consider the plausibility of the power-law fit, we see fewer networks. 62\% of the preferential attachment graphs fall into the Weakest and Weak categories, 60\% in the Strong category, and 0 in the Strongest category. This is not entirely unexpected, however. Each of these graphs is directed and splits into three degree sequences: one each for in-degree, out-degree, and total degree. As predicted by the literature, the in-degree sequences are usually plausibly power-law (80\%), while the out-degree sequences never are, and the total degree sequences are again usually but not always power-law (74\%). This accounts for the variance and lower-percentage of networks showing direct power-law evidence. This also implies that the preferential attachment mechanism does better at generating networks for which a power law is a better fit than alternatives, than at generating genuinely power-law networks. The absence of graphs in the Strongest category is due to the fact that this category requires that 90\% of associated simple graphs be plausibly power-law, and this generating mechanism gives at best 67\%.

The vertex-copying method \cite{Newman2010} is also expected to generate scale free structure. We again start with a 4-node directed network, each node connected to the other 3. We add nodes one at a time until we reach 5000 total and for each follow the same procedure. Pick an existing node $u$ at random. For each out-edge that $u$ has, copy it to the new node $v$ with probability $q=0.6$. With probability $1-q=0.4$, attach uniformly to one of the other nodes. Each node has out-degree 3. When we process this graph we expect again a power-law in-degree distribution and usually a power-law total-degree distribution.

The results are consistent with what we expect. 88\% of the vertex-copying graphs are Super-Weakly scale-free, which jumps to 99\% if we ignore power-law with cutoff. 74\% fall into the Weakest and Weak categories, meaning the power law is plausible with at least 50 points in the tail of the degree sequence. 70\% fall into the Strong category. Since out-degree sequences are never plausibly power-law we have none in the Strongest category.

Erd\H{o}s-R\'enyi random graphs are simple and are generally known to have thinner-tailed degree distributions. To generate these graphs, we add $n=5000$ nodes and then for every possible edge, there is a probability $p=c/(n-1)$ of connection, where $c$ is the desired mean degree. We chose $c=6$, so $p=6/4999\approx 0.0012$. We expect the majority of these to follow thinner-tailed distributions than the power law.

The Erd\H{o}s-R\'enyi random networks indeed have very different results. Only 16\% are Super-Weak, though this increases to 31\% without the power law with cutoff as an alternative. This indicates that these networks have thinner-tailed degree distributions. 51\% and 50\% of these networks fall into the Weakest and Weak categories, respectively, but because the best-fit $\alpha$-values are all large (the smallest is 7.55), none fall into the Strong or Strongest categories. This is again consistent with the thin-tailed nature of these networks. We may have expected to see a smaller fraction of these networks falling into the Weak categories, but this result indicates how inclusive these two categories are.

As our methods were able to fairly well recover the ground truth when tested on these synthetic networks, we feel confident in our results for the real-world networks.

\end{document}